\documentclass[aps,twocolumn,floatfix, showkeys, superscriptaddress, nofootinbib]{revtex4}
\usepackage{amsmath}
\usepackage{graphicx,bm,hhline}
\usepackage{calrsfs}
\DeclareMathAlphabet{\pazocal}{OMS}{zplm}{m}{n}

\newcommand{\bcri}{{\bm R}_i}

\newcommand{\bq}{{\bm q}}

\newcommand{\br}{{\bm r}}
\newcommand{\brp}{{\br}^\prime}

\newcommand{\hf}{\frac{1}{2}}

\newcommand{\wig}[1]{\mathrel{\hbox{\hbox to 0pt{\lower.6ex\hbox{$\sim$}\hss}\raise.4ex\hbox{$#1$}}}}

\begin{document}
\title{Potential of mean force for electrical conductivity of dense plasmas}

\author{C. E. Starrett}
\email{starrett@lanl.gov}
\affiliation{Los Alamos National Laboratory, P.O. Box 1663, Los Alamos, NM 87545, U.S.A.}


\date{\today}
\begin{abstract}
The electrical conductivity in dense plasmas can be calculated with the relaxation-time
approximation provided that the interaction potential between the scattering electron
and the ion is known.  To date there has been considerable uncertainty as to the
best way to define this interaction potential so that it correctly includes
the effects of ionic structure, screening by electrons and partial ionization.  Current approximations
lead to significantly different results with varying levels of agreement when compared to 
bench-mark calculations and experiments.  We present a new way to define this
potential, drawing on ideas from classical fluid theory to define a potential
of mean force.  This new potential results in significantly improved agreement with
experiments and bench mark calculations, and includes all the aforementioned
physics self-consistently.
\end{abstract}
\pacs{ }
\maketitle

\section{Introduction}
Accurate prediction of electronic conductivity in dense plasmas has proved
to be a challenging problem.  One approach is to use density functional theory molecular
dynamics (DFT-MD) coupled with the
Kubo-Greenwood formalism \cite{desjarlais02, desjarlais17, sjostrom15, holst11, french14, hu14}.
This method is thought to be accurate but is 
limited in the problems it can be applied to due to its significant computational
expense.  Moreover, fundamental questions like its treatment of electron-electron
collisional effects are still active areas of research \cite{desjarlais17}.
Experimentally, the subject is an active area of research \cite{sperling15} and
has a long history \cite{milchberg88, clerouin12, clerouin08, desilva98, benage99},
in part due to the difficulty in obtaining model independent measurements.

Another class of methods starts from the Boltzman equation and introduces
a relaxation-time approximation \cite{lee84} in which electrons are scattered 
from ions and other electrons.  The question then becomes one of calculating
the electron-ion and electron-electron cross sections.  In the degenerate electron limit
this approach becomes identical to the famous Ziman method \cite{ziman61,burrill16}.

To calculate the electron-ion cross section in dense plasmas
one needs to be able to model partial ionization induced by density and
temperature, and also the corresponding changes in the ionic structure.
Average atom models have long been used for this purpose 
\cite{rinker88, scaalp, sterne07, ovechkin16, burrill16, perrot87, starrett12a}.
They are DFT based models that attempt to calculate the properties of
one averaged atom in the plasma.  They are computationally efficient and
are well suited to making wide ranging tables of data \cite{rinker88}.

While on the face of it, it seems natural to couple these average atoms 
to the relaxation time approximation and
thus calculate conductivities, it turns out that the results are sensitive
to exactly how this coupling is done. Specifically, one needs to 
define an electron-ion scattering potential, and how this definition
is made strongly effects the resulting conductivities 
\cite{ovechkin16, burrill16, starrett16b, perrot99}.

In this paper we explore a new method of coupling average atom models to the 
relaxation-time approximation that extends the potential
of mean force from classical fluid theory \cite{baalrud13}
to the quantum domain.  Thus, a new quantum potential of mean force is defined.
It includes correlations with electrons and ions
surrounding the central scatterer through the quantum fluid
equations known as the quantum Ornstein-Zernike equations \cite{chihara91}.
We find that the new potential leads to generally more accurate conductivity predictions when
to compared to DFT-MD simulations and experiments for aluminum over
a wide range of conditions.
We also explore the influence of electron-electron collisions by
including a correction factor due to Reinholz {\it et al} \cite{reinholz15}, and find
that plays a significant and important role in certain cases.

\section{Theoretical model\label{sec_mod}}
\subsection{Conductivity in terms of the electron relaxation time}
In the relaxation-time approximation the conductivity is given by
\footnote{Unless otherwise stated, atomic units are used throughout in
which $\hbar = m_e = e = k_B = a_B = 1$, and the symbols have their usual
meanings.}
\begin{equation}
\sigma_{DC} = \int_0^\infty \left( -\frac{df}{d\epsilon} \right)  N_e(\epsilon) \tau_\epsilon d\epsilon
\end{equation}
where $\tau_\epsilon$ is the energy dependent relaxation time,  $f$ is the Fermi-Dirac occupation factor,
\begin{equation}
N_e(\epsilon) = n_I^0 \int_0^\epsilon d\epsilon^\prime \chi(\epsilon^\prime) 
\label{nee}
\end{equation}
and $\chi(\epsilon)$ is the density of states such that the number of valence electrons per atom
$\bar{Z}$ is
\begin{equation}
\bar{Z} = \frac{\bar{n}_e^0}{n_I^0} = \int_0^\infty d\epsilon \chi(\epsilon)  f(\epsilon, \mu_e)
\label{zbar}
\end{equation}
Here $\mu_e$ is chemical potential and $\bar{n}_e^0$ ($n_I^0$) is the density of
valence electrons (ions).  
If we take the density of states to be its free electron form
\begin{equation}
\chi^{free}(\epsilon) = \frac{\sqrt{2 \epsilon}}{n_I^0 \pi^2 }
\end{equation}
then
\begin{equation}
N_e^{free}(\epsilon) = \frac{ v^3 }{3 \pi^2 }
\end{equation}
where $\epsilon=v^2/2$.  The resulting equation that we will use here is
\begin{equation}
\sigma_{DC} = \frac{1}{3\pi^2} \int_0^\infty \left( -\frac{df}{d\epsilon} \right)  v^3 \tau_\epsilon d\epsilon
\end{equation}

\subsection{The electron relaxation time's relation to the momentum transport cross section}
The relaxation time is related to the mean free path $\lambda_\epsilon$
by
\begin{equation}
\tau_\epsilon = \frac{\lambda_\epsilon} {v}
\end{equation}
which is in turn related to the momentum transport cross section $\sigma_{TR}(\epsilon)$
\begin{equation}
\lambda_\epsilon = \frac{1}{n_I^0 \sigma_{TR}(\epsilon) }
\end{equation}
hence
\begin{equation}
\tau_\epsilon = \frac{1}{n_I^0\,v\, \sigma_{TR}(\epsilon) }
\end{equation}

\subsection{The momentum transport cross section in terms of scattering phase shifts}
The momentum transport cross section is calculated from 
\begin{equation}
\sigma_{TR}(p) = 2\pi \int_0^\pi d\theta 
\frac{d\sigma}{d\theta}(\epsilon,\theta)
(1-\cos\theta)\sin\theta
\label{tr}
\end{equation}
where $p=m_e v$ and $\frac{d\sigma}{d\theta}(\epsilon,\theta)$ is the differential cross section for one
scatterer in the plasma.  This can be calculated from the scattering phase shifts $\eta_l(\epsilon)$
once the single center scattering potential $V^{scatt}(r)$ is known.
\begin{equation}
\frac{d\sigma}{d\theta}(\epsilon,\theta) = \left| \pazocal{F}(\epsilon,\theta)\right|^2
\label{dsdt}
\end{equation}
where the scattering amplitude $\pazocal{F}$ is
\begin{equation}
\pazocal{F}(\epsilon,\theta)
 = \frac{1}{p}
\sum\limits_{l=0}^{\infty} (2l+1) \sin \eta_l e^{\imath \eta_l} P_l(\cos\theta)
\label{scattamp}
\end{equation}
The angular integral in equation (\ref{tr}) can be carried out analytically, yielding
\begin{equation}
\sigma_{TR}(\epsilon) = \frac{4 \pi}{p^2} \sum\limits_{l=0}^{\infty} (l+1) \left( \sin\left( \eta_{l+1} - \eta_{l} \right)\right)^2
\label{tr2}
\end{equation}


\subsection{Choice of scattering potential}
To calculate the scattering amplitude $\pazocal{F}$ or momentum transfer cross section 
$\sigma_{TR}(r)$ we need a scattering potential $V^{scatt}(r)$.  The physical
picture inherent to our application of the relaxation time approximation is that electrons 
scatter off one center at a time, and that the plasma
is made up of identical scattering centers.  The question then becomes: what defines a scattering
center?  
\subsubsection{Pseudoatom (PA) potential}
Lets assume that the plasma is made up of an ensemble of identical pseudoatoms.  Each pseudoatom
is a nucleus and a spherically symmetric screening cloud of electrons with density $n_e^{PA}(r)$.  
The total potential for the plasma is then
\begin{equation}
V({\bm r}) = \sum_{i=1}^N V^{PA}(|\br -\bcri|)
\label{v}
\end{equation}
where the sum is over the $N$ nuclei in the plasma and
\begin{equation}
V^{PA}(r) = -\frac{Z}{r} + \int d^3r^\prime \frac{n_e^{PA}(r^\prime)}{\left| \br - \brp \right|}
+ V^{xc}[n_e^{PA}(r)]
\end{equation}
with $V^{xc}[n]$ being the contribution from electron exchange and correlation.

In the high energy limit the Born cross section becomes accurate and the scattering cross
section for the whole plasma is
\begin{equation}
\frac{d\sigma^{plasma}}{d\theta}(\epsilon,\theta) = \left| \frac{V({\bm q})}{2\pi} \right|^2
\end{equation}
where
\begin{equation}
q^2 = 2p^2[1-\cos\theta]
\end{equation}
and the Fourier transform of the potential is 
\begin{equation}
V({\bm q}) = \int d^3r e^{-\imath \br\cdot\bq} V(\br)
\end{equation}
Taking the Fourier Transform of equation (\ref{v}) and using the definition of the ionic structure factor
\begin{equation}
S_{\rm ii}(q) = \frac{1}{N}\left< \rho_{\bq} \rho_{-\bq} \right>
\label{sii_micro}
\end{equation}
where
\begin{equation}
\rho_{\bm q} = \sum_i e^{-\imath \bq \cdot \bcri }
\label{micro}
\end{equation}
and the angular brackets indicate that the configurational average
has been taken, the differential cross section per scattering center becomes
\begin{equation}
\frac{d\sigma}{d\theta}(\epsilon,\theta) = S_{ii}(q) \left| \frac{V^{PA}({q})}{2\pi} \right|^2
\label{dsdt_pa}
\end{equation}
This is valid when the Born approximation is accurate (i.e. for weak scattering, typically high energy scattering electrons).
To return to the strong scatterer picture for which the Born approximation is invalid, one
approach \cite{burrill16} is to replace the Born cross section $|V^{PA}(q)/2\pi|^2$ with its t-matrix equivalent, equations
(\ref{dsdt}) and (\ref{scattamp}), i.e.
\begin{equation}
\frac{d\sigma}{d\theta}(\epsilon,\theta) = S_{ii}(q) \left| \pazocal{F}^{PA}(\epsilon,\theta)\right|^2
\label{dsdt_tpa}
\end{equation}
where $\pazocal{F}^{PA}$ is the scattering amplitude for the $V^{PA}(r)$ potential.
However, as the differential cross section now depends on 
$S_{ii}(q)$, the angular integral in equation (\ref{tr}) can longer be done analytically and must be done numerically
\footnote{From a numerical point of view the angular integral in equation (\ref{tr}) is not particularly difficult.}.

Using this approach the Born limit of the scattering cross section is recovered, and so the method
should be accurate at high temperatures or high densities where the scattering
electrons have high energies.  However, at relatively lower densities and temperatures (like room temperature
and pressure), the Born cross section will be significantly in error, and therefore equation (\ref{dsdt_tpa}) may also be significantly
in error.

\subsubsection{Average atom (AA) potential}
An alternative and routinely used \cite{ovechkin16,faussurier15,pain10, rinker88, sterne07, rozsnyai08, starrett12a} definition of the scattering potential is to use an average atom potential $V^{AA}(r)$.  There
are a number of realistic variations on how this should be defined but such variations are
relatively unimportant for present purposes.  We define
\begin{equation}
V^{AA}(r) = -\frac{Z}{r} + \int_{r<R_{WS}} d^3r^\prime \frac{n_e^{AA}(r^\prime)}{\left| \br - \brp \right|}
+ V^{xc}[n_e^{AA}(r)]
\end{equation}
where the integral and the potential are confined to the ion sphere with (Wigner-Seitz) radius $R_{WS}$.
The ion-sphere is required to the charge neutral and typically $V^{AA}(r) = 0$ for $r>R_{WS}$.
In this approach
\begin{equation}
\frac{d\sigma}{d\theta}(\epsilon,\theta) = S_{ii}(q) \left| \pazocal{F}^{AA}(\epsilon,\theta)\right|^2
\label{dsdt_aa}
\end{equation}
In deriving this result \cite{evans73} one assumes a muffin-tin potential and a crude estimate
of multiple scattering effects gives rise to the ionic structure factor $S_{ii}(q)$.
The average atom potential $V^{AA}(r)$ cannot recover the Born limit and equation (\ref{dsdt_aa}) 
will be incorrect in the high temperature or high density limit.

\begin{figure}
\begin{center}
\includegraphics[scale=0.40]{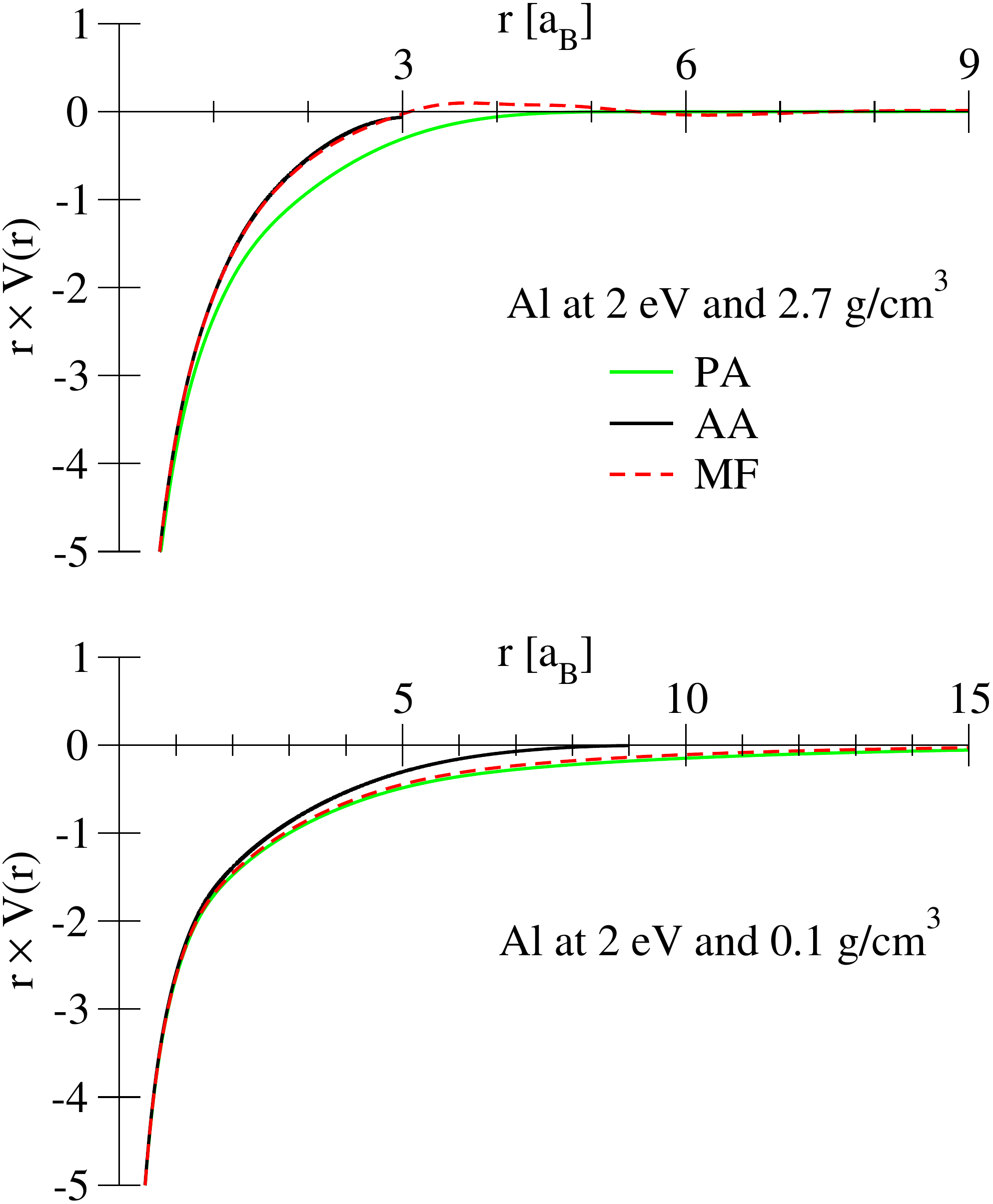}
\end{center}
\caption{(Color online) Examples of scattering potentials for two different
aluminum cases.  In the top panel, for aluminum at 2.7 g/cm$^3$ and 2 eV,
 the mean force potential is close to the
average atom potential (which only extends to the ion sphere radius).
In the bottom panel, for aluminum at 0.1 g/cm$^3$ and 2 eV, the mean force
potential is closer to the pseudoatom potential.
}
\label{potentials}
\end{figure}

\subsubsection{Potential of mean force (MF)}
We desire the potential felt by one electron as it scatters from one center.  This scattering 
does not happen in isolation as in the $V^{PA}(r)$ approximation (equation (\ref{dsdt_tpa})).  
For classical (``$cl$'') particles the Ornstein-Zernike
equations can be used to define a potential of mean force
that includes the potential created by the central scatterer as well as its correlations with surrounding 
scattering centers.  In the hyper-netted chain closure approximation (HNC) this reads
\begin{eqnarray}
V^{MF,cl}(r) & = & V(r) - \frac{1}{\beta} \left( h(r) - C(r)\right)\nonumber\\ 
                    & = & V(r) + n^0 \int d^3 r^\prime \frac{ C(|\br -\brp|) }{-\beta} h(r^\prime) 
\end{eqnarray}
where $h(r)$ is the pair correlation function\footnote{$h(r)$ is simply related the structure factor $S(q) = 1+n^0 h(q)$.}, 
$C(r)$ the direct correlation function, $n^0$ is the particle
density, $V(r)$ the direct interaction potential between two particles (e.g. the coulomb potential), and $\beta$ is the inverse
temperature.  Such a 
mean field potential has been used successfully for calculation of ionic transport quantities \cite{daligault16, baalrud13}.  The
analogous electron-ion potential when considering quantal electrons is given by the quantum Ornstein-Zernike
equations as \cite{chihara91, starrett14b}
\begin{eqnarray}
V^{MF}(r) & = & V_{ie}(r) + n_i^0 \int d^3 r^\prime \frac{ C_{ie}(|\br -\brp|) }{-\beta} h_{ii}(r^\prime) \nonumber\\
               &   & + \bar{n}_e^0 \int d^3 r^\prime \frac{ C_{ee}(|\br -\brp|) }{-\beta} h_{ie}(r^\prime) \nonumber\\
\label{vmf1}               
\end{eqnarray}
where $C_{ie}$ ($C_{ee}$) is the electron-ion (-electron) direct correlation function.
The quantum Ornstein-Zernike equations are valid when the electrons respond linearly to the ions, i.e. in the linear
response regime.  Thus is it necessary to artificially separate the electrons into two groups: those that are 
bound the nucleus ($n_e^{ion}(r)$) and therefore respond very non-linearly, and those that are not bound ($n_e^{scr}(r)$), and for which
linear response is reasonably accurate.  This was the approach taken in \cite{starrett13,starrett14} where
the quantum Ornstein-Zernike equations where solved and resulting pair distribution functions where
shown to be accurate.  Adopting that model, $V_{ie}(r)$ becomes
\begin{equation}
V_{ie}(r) = -\frac{Z}{r} + \int d^3r^\prime \frac{n_e^{ion}(r^\prime)}{\left| \br - \brp \right|}
+ V^{xc}[{n_e^{ion}}(r)]
\end{equation}
Using the relation \cite{starrett13}
\begin{equation}
C_{ij}(k) = -\beta V_{ij}^C(k) + \widetilde{C}_{ij}(k)\label{cij_tilde}
\end{equation}
where $V_{ij}^C(k) = Z_i Z_j 4\pi/k^2$ is the Coulomb potential in Fourier Space between two charges $Z_i$ and $Z_j$,
and collecting terms gives
\begin{eqnarray}
V^{MF}(r) & = & V^{PA}(r) + n_i^0 \int d^3 r^\prime \frac{ -\bar{Z} h_{ii}(r^\prime) + n_e^{ext}(r^\prime)}{|\br -\brp|} \nonumber\\
          &   & + V^{xc}[n_e^{ext}(r)] \nonumber\\
          &   & + n_i^0 \int d^3 r^\prime \frac{ \widetilde{C}_{ie}(|\br -\brp|) }{-\beta} h_{ii}(r^\prime)
\end{eqnarray}
with
\begin{eqnarray}
n^{ext}_e(r) & = & \int d^3 r^\prime n_e^{scr}(r^\prime) h_{ii}(|\br -\brp|) 
\end{eqnarray}
and $\bar{Z} = \bar{n}_e^0 / n_i^0 = \int d^3r\, n_e^{scr}(r)$.
$V^{MF}(r)$ is the scattering potential for one scattering center, and implicitly includes the ionic
structure ($S_{ii}(q) = 1 + n_i^0  h_{ii}(q)$), i.e. $S_{ii}(q)$ does not explicity appear as in equations (\ref{dsdt_tpa}) and (\ref{dsdt_aa}) for
the pseudoatom and average atom differential cross sections.  Hence, the angular integral in equation (\ref{tr}) can be carried out
analytically and we directly solve equation (\ref{tr2}).
\begin{figure}
\begin{center}
\includegraphics[scale=0.30]{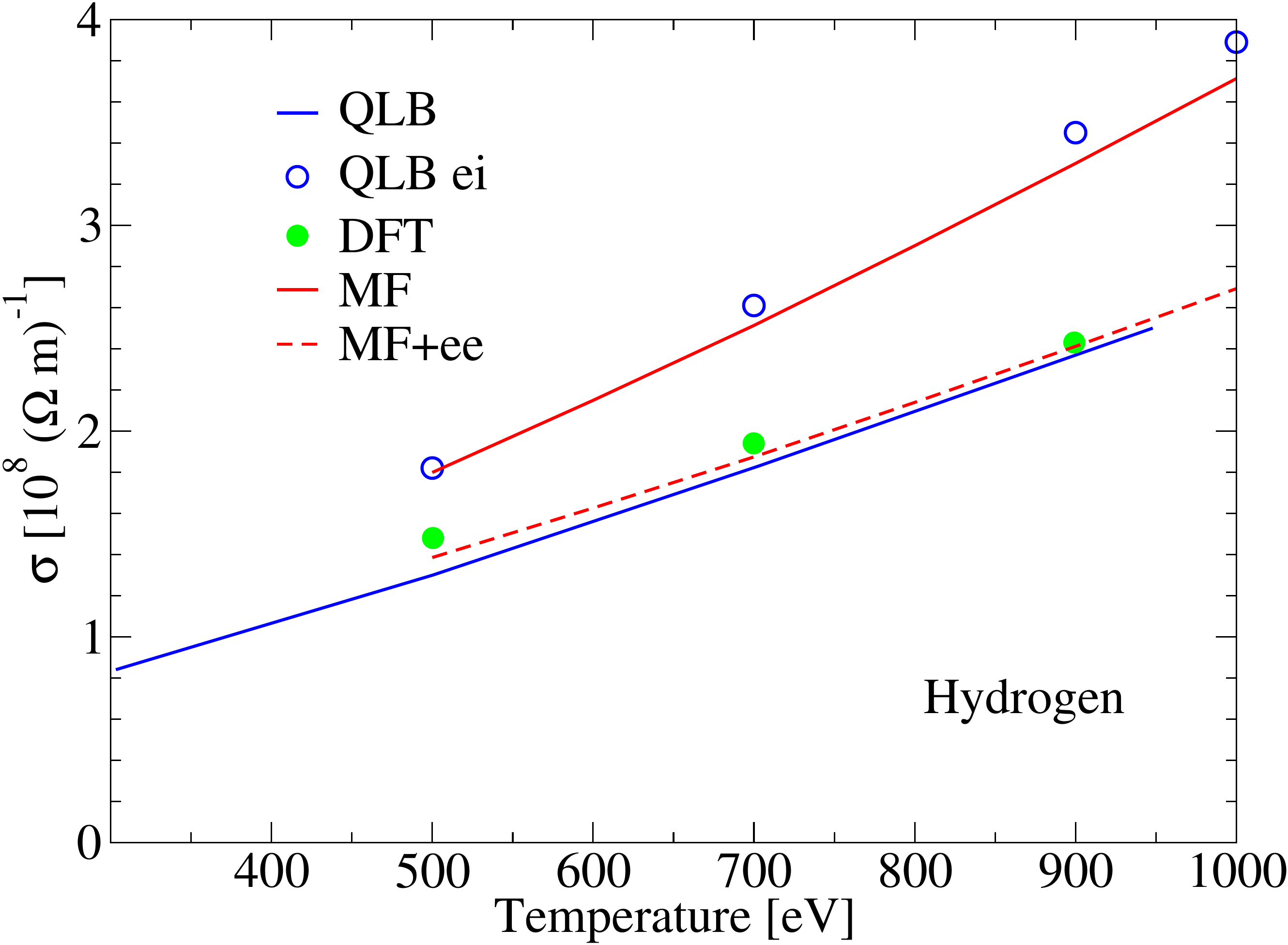}
\end{center}
\caption{(Color online) The effect of electron-electron collisions on the electrical
conductivity of hydrogen at 40 g/cm$^3$.  Compare with figure 3a of reference \cite{desjarlais17}.
Our calculations using the mean force potential (MF) agree well with the quantum Lenard-Balescu calculations
with only electron-ion collisions included (QLB ei).  Including a correction factor that accounts for
electron-electron collisions due to Reinholz {\it et al} \cite{reinholz15}, our calculations
(MF+ee) fall into close agreement with the QLB calculation that include electron-electron collisions,
as well as DFT simulation results.
}
\label{ee}
\end{figure}
\begin{figure}
\begin{center}
\includegraphics[scale=0.40]{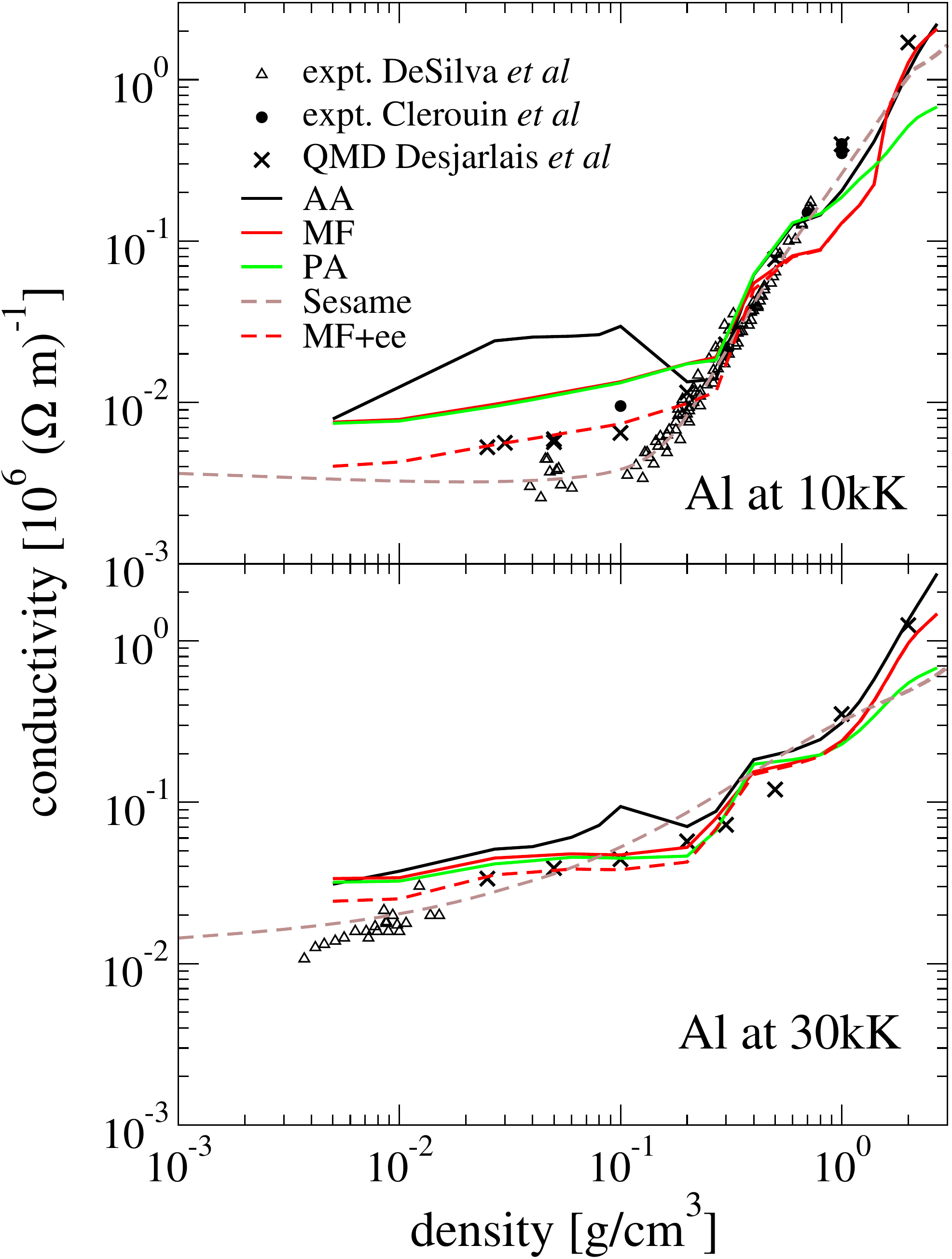}
\end{center}
\caption{(Color online) 
Electrical conductivity of aluminum at 10 kK (top panel) and 30 kK (bottom panel) as calculated in the 
present approach for the three potentials  $V^{AA}(r)$, $V^{PA}(r)$ and $V^{MF}(r)$.  Comparisons
are made to QMD simulations of Desjarlais et al \cite{desjarlais02} and to the experiments
of DeSilva {\it et al} \cite{desilva98} and Cl\'erouin {\it et al} \cite{clerouin08}.  MF+ee
refers to calculation using $V^{MF}(r)$ and explicitly accounting for electron-electron
collisions using the fit formula of reference \cite{reinholz15}.  
}
\label{fig_des}
\end{figure}

\begin{figure}
\begin{center}
\includegraphics[scale=0.4]{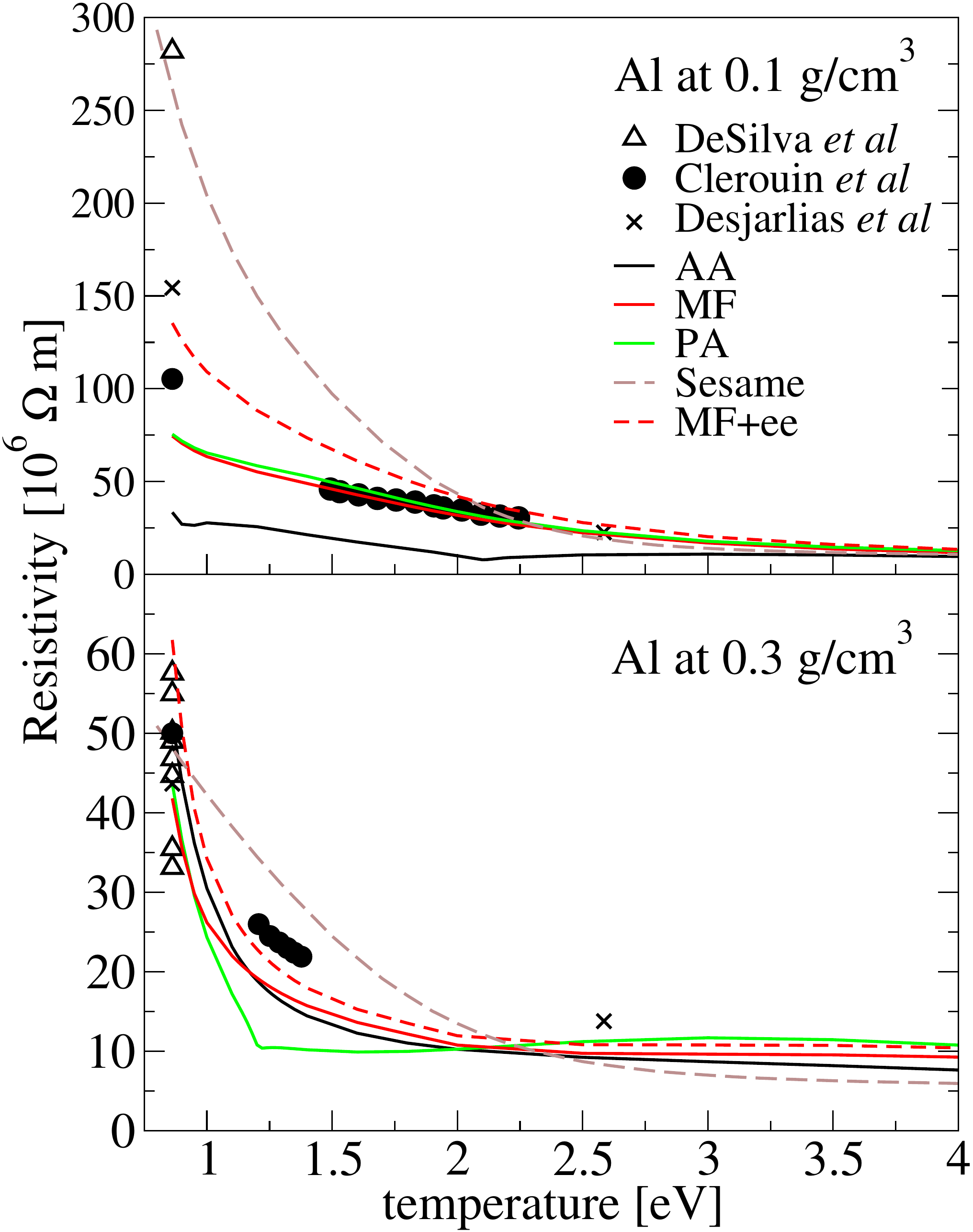}
\end{center}
\caption{(Color online) Electrical resistivity of aluminum at 0.1 g/cm$^3$ (top panel) and
0.3 g/cm$^3$ (bottom panel).   Experiments are from Cl\'erouin {\it el at }\cite{clerouin12} and
DeSilva {\it et al} \cite{desilva98}.  Also shown are DFT-MD (QMD) results of Desjarlais {\it et al} \cite{desjarlais02}.
Present results use the three potentials discussed in the text: $V^{AA}(r)$, $V^{PA}(r)$ and $V^{MF}(r)$,
as well as the effect of electron-electron collisions for $V^{MF}(r)$ (MF+ee).
Also shown is Sesame 29371 which is based on the models of references \cite{desjarlais01,lee84}.
}
\label{fig_cl}
\end{figure}

In the high temperature regime $h_{ii}(r) \to 0$ $\forall r$, and $\widetilde{C}_{ie}(r) \to 0$ $\forall r$,
hence $V^{MF}(r) \to V^{PA}(r)$, and the Born limit will be recovered.  

It is interesting to note that the potential of mean force obtained above (equation (\ref{vmf1})) is that same
as the potential used in Chihara's QHNC model \cite{chihara91}.  There it has not been used for conductivity calculations.

In figure \ref{potentials} examples of these three potentials $V^{AA}(r)$, $V^{MF}(r)$ and $V^{PA}(r)$
are shown for aluminum at 2 eV and 2.7 g/cm$^3$ (top panel) and 0.1 g/cm$^3$ (bottom panel).  For the higher density
case correlations are important and $V^{MF}(r)$ is close to $V^{AA}(r)$, whereas for the lower density case
correlations are less important and $V^{MF}(r)$ is closer to $V^{PA}(r)$.

\section{Numerical results}

To generate the potentials needed for calculation of the conductivities ($V^{AA}(r)$, $V^{PA}(r)$ and $V^{MF}(r)$) we have used the model of references 
\cite{starrett13, starrett14}.  In this model DFT is used to determine the electronic structure of one average
atom in the plasma.  This provides a closure relation for the quantum Ornstein Zernike equations,
which are thus solved self-consistently for all quantities of interest (eg. $S_{ii}(k)$, $C_{ie}(k)$, $\bar{n}_e^0$ etc.)
This model has been shown to be realistic for equation of state \cite{starrett16} and
ionic structure \cite{starrett15b}.  It is a plasma model, and as such is most accurate at elevated temperatures.  For example,
for aluminum at solid density realistic structure factors were predicted for temperatures greater than $\sim$1 eV \cite{starrett13}.
A numerical issue for the solution of equation (\ref{tr2}) at very high temperatures is discussed in the appendix.

The conductivity model discussed in section \ref{sec_mod} includes electron collisions with ions in the plasma.  There is no explicit account
of electron-electron collisions.  This difficult problem has been discussed by a number
of authors \cite{desjarlais17, reinholz15, potekhin96, kuhlbrodt01, wardana06}.  Here we
test the effect of including electron-electron collisions by using the fit formula
due to Reinholz {\it et al} \cite{reinholz15}.  This formula takes as input
the ion density, temperature and the average ionization (which is provided by
the model of references \cite{starrett13, starrett14}).  Very recently, the
effect of electron-electron collisions on electrical conductivity was considered in
detail \cite{desjarlais17} by comparing quantum Lenard-Balescu calculations
that explicitly account for both electron-electron and electron-ion collisions, to
DFT-MD simulations that use the Kubo-Greenwood formalism. In figure \ref{ee}
we compare to the results of reference \cite{desjarlais17} for hot, dense hydrogen.
From the figure it is clear that including only electron-ion collisions in the
present method leads to good agreement with the QLB results that also only include
electron-ion collisions.  Adding the electron-electron collisions factor to the mean force
results, we now see agreement with the QLB collision that also explicitly account for these,
as well as good agreement with the DFT results.  This comparison strongly indicates
that it is necessary to explicitly account of electron-electron collisions in our
relaxation time approach.

In figure \ref{fig_des} we compare to the experiments of references \cite{desilva98} and
\cite{clerouin08} for aluminum at two temperatures (10 kK and 30 kK) as a function
of density.  We also show DFT-MD (also known as QMD) results from reference \cite{desjarlais02},
which is a less approximate method than the present, but is much more computationally expensive.
Results using $V^{AA}(r)$ are in reasonable agreement with the QMD calculations at high densities, but
tend to be too large at lower densities, compared to the experimental data.  In contrast,
$V^{PA}(r)$ gives reasonable results at lower densities, but underestimates the QMD results
at high densities.  $V^{MF}(r)$ gives reasonable, but not perfect, agreement at both low and
high densities, for both temperatures.  Including the electron-electron collision factor, agreement
with the experiment and the DFT-MD simulations for the mean force potential is markedly improved.   
This further strengthens the case that electron-electron collisions should be explicitly accounted for
when using the relaxation time approach.  We note that the correction factor is the same
for all three potentials, but its effect is only shown for $V^{MF}(r)$ for clarity.
\begin{figure}
\begin{center}
\includegraphics[scale=0.3]{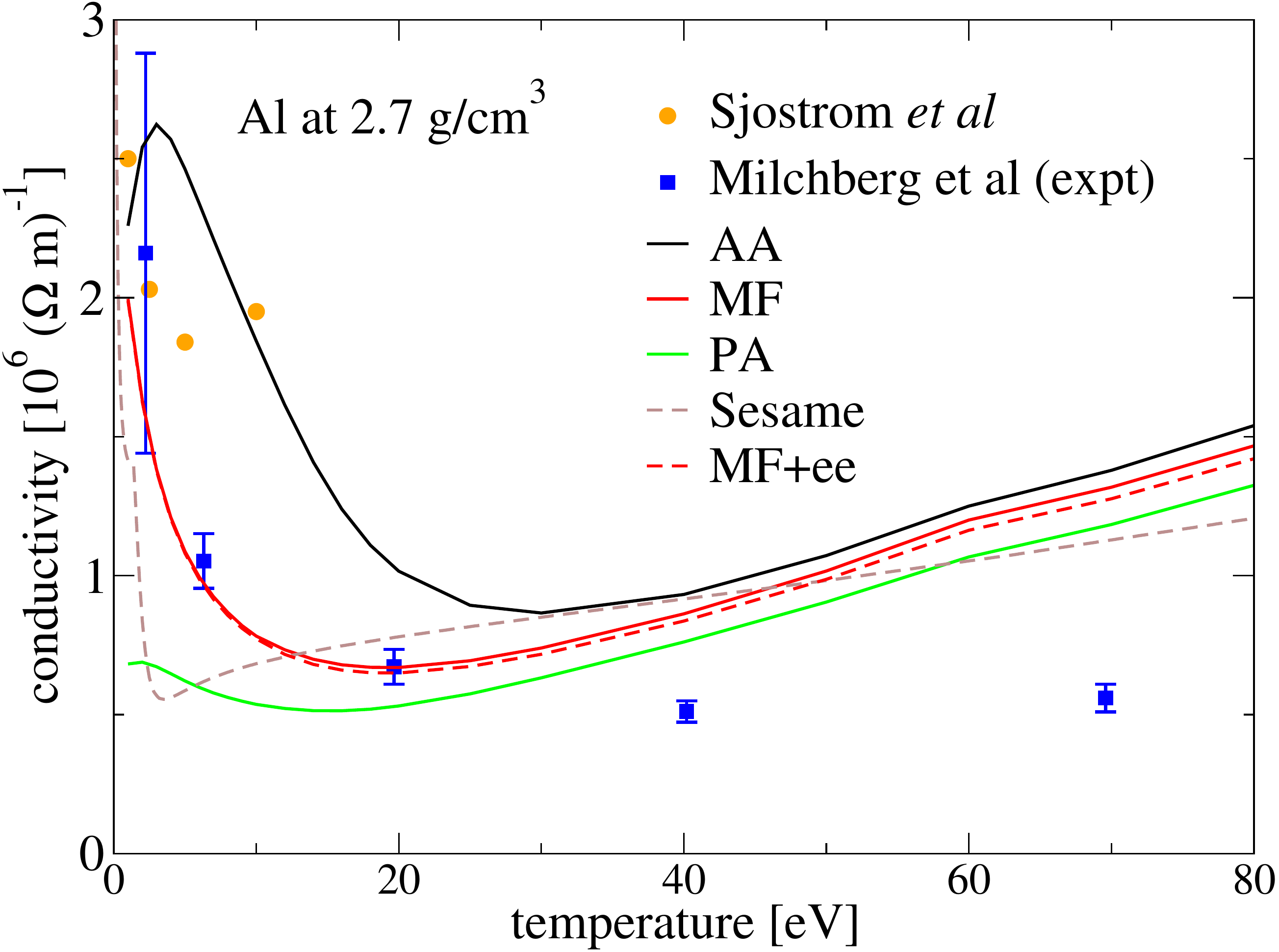}
\end{center}
\caption{(Color online) 
Electrical conductivity of aluminum at 2.7 g/cm$^3$ as calculated in the
present approach for the three potentials  $V^{AA}(r)$, $V^{PA}(r)$ and $V^{MF}(r)$, compared 
to the experiment of Milchberg {\it et al} \cite{milchberg88}.  Also shown are DFT-MD
results due to Sjostrom {\it et al} \cite{sjostrom15} and Sesame 29371.
}
\label{fig_mil}
\end{figure}

It should also be noted that the range of conditions plotted in figure \ref{fig_des} is a particularly
challenging regime to model due to the relocalization of electrons as density is lowered.
This relocalization explains the structures in the present results, which are a consequence
of the spherical symmetry in the underlying model \cite{starrett13,starrett14}, and
are probably too pronounced.  
Also shown in the figure are Sesame 29371 curves, which were designed to fit the experiments \cite{desilva98} and as a consequence
agree very well with those experiments.  At the highest densities shown, the Sesame
curve underestimates the QMD calculations, particularly at 30 kK.

In figure \ref{fig_cl} calculations of the resistivity
of warm dense aluminum at 0.1 g/cm$^3$ and 0.3 g/cm$^3$  are compared
to the experimental results of Cl\'erouin {\it et al} \cite{clerouin12, clerouin08} 
and DeSilva {\it et al} \cite{desilva98} as well as DFT-MD results from reference \cite{desjarlais02}.
Results using all three types of potential are shown.  
All the theory curves are reasonably close to each other for temperatures
greater than 2 eV. For temperatures lower that this the variance significantly increases.  
Compared to the Cl\'erouin {\it et al} data the mean-force potential gives reasonable agreement 
for both densities, whereas neither $V^{AA}(r)$ or $V^{PA}(r)$
give a consistent level of agreement with the data at both densities.  $V^{AA}(r)$
is closer to the data at the higher density, where correlations with surrounding ions are relatively important,
while $V^{PA}(r)$ is closer at the lower density, where the influence of correlations
reduced.  While the agreement of the $V^{MF}$ calculations is not perfect it should
be remembered that neither is the underlying model \cite{starrett13,starrett14} or
the experiments, which themselves do not agree with each other.  
The effect of electron-electron collisions is significant in for the lower temperatures.
We note that the temperatures of the experiments of Cl\'erouin {\it et al} are obtained through comparison
with DFT-MD simulations \cite{clerouin12} and are not measured directly.  
For completeness we have also compared
to Sesame calculations that are based on the model of \cite{desjarlais01}.  The 
Sesame result tends to overestimate the Cl\'erouin {\it et al} data.  They are fit
to the DeSilva {\it et al} data \cite{desjarlais01} and are therefore is good agreement with those
experiments.  
\begin{figure}[t!]
\begin{center}
\includegraphics[scale=0.3]{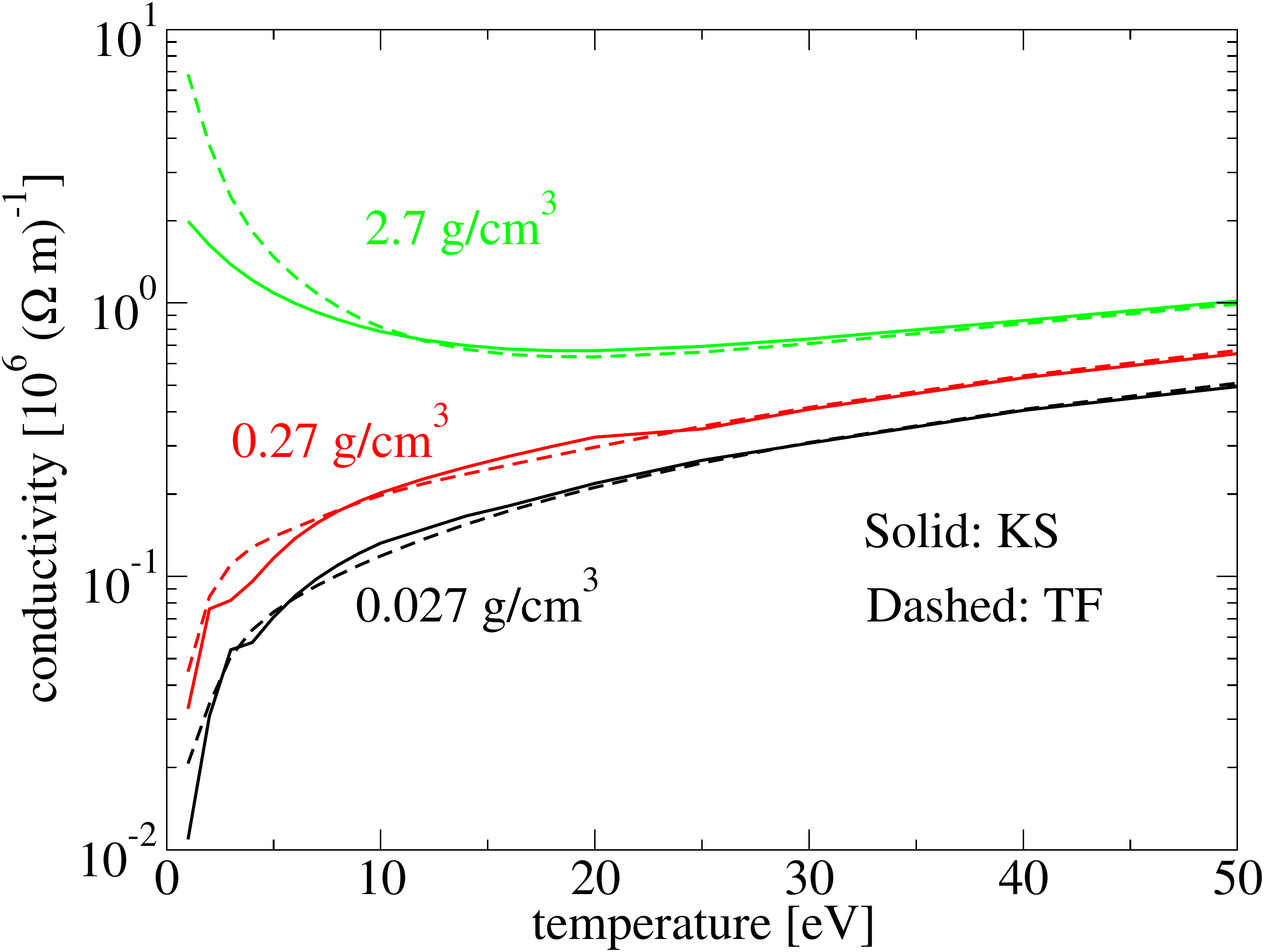}
\end{center}
\caption{(Color online) 
Electrical conductivity of aluminum using the $V^{MF}(r)$ potential (no electron-electron factor).  
Results using a potential created by a Kohn-Sham calculation are compared to those created using a Thomas-Fermi
calculation for three densities.  For the cases shown the two calculations agree
well for temperatures greater than roughly 25 eV.
}
\label{fig_tf}
\end{figure}

In figure \ref{fig_mil} we compare to the experiment of reference \cite{milchberg88}.  This
experiment reported results for solid density aluminum up to high temperatures.  We also
compare to DFT-MD results from reference \cite{sjostrom15}.  At the lower temperatures (20 eV and below) 
the results using $V^{MF}(r)$ agree well with the experiment.  They underestimate the
DFT-MD results of \cite{sjostrom15} by $\sim 20\%$ at 1 eV.  Note that the increase in conductivity
in going from 5 to 10 eV reported by Sjostrom {\it et al} \cite{sjostrom15} may be an artifact
of the pseudopotential used \cite{sjostrom_p}.  The $V^{PA}(r)$ results
significantly underestimate the conductivity in the low temperature regime compared
to both the DFT-MD and experimental results.  The $V^{AA}(r)$ results over predict the
conductivity in this regime.  At higher temperatures none of the results from the potentials match the
experiment.  However, it has been pointed out \cite{dwp92} that the temperatures
reported in \cite{milchberg88} are dependent on a model for ionization, and that
different models yield significantly different temperatures (eg. the 40 eV point
becomes 23.5eV).  Moreover, while dc conductivity was reported, it as an ac conductivity 
at 4.026 eV that was measured.  Finally, we also show the Sesame result.  This curve
has a much sharper drop at low temperatures, and a different slope as temperature is
increased.  Over the temperature range shown the maximum difference between Sesame and the
$V^{MF}(r)$ curve is a factor of $\sim 2$, but is generally closer.  The effect of
electron-electron collisions is small due to the relatively high degeneracy of the plasma
and relatively large electron-ion collision cross section.

While generation of the potential $V^{MF}(r)$ using the model of references \cite{starrett13,starrett14}
is computationally inexpensive relative to DFT-MD simulations, is still takes 30 minutes to 1 hour per
density temperature point.  With a view to generating tables of conductivity data it is desirable to find
an even more computationally efficient model.  One option is to use a Thomas-Fermi based DFT model, in place
of the Kohn-Sham description used above \cite{starrett13,starrett14}.  With such a model it takes seconds to generate 
$V^{MF}(r)$.  In figure \ref{fig_tf} we compare conductivities using the Kohn-Sham and Thomas-Fermi methods of generating 
$V^{MF}(r)$.  We find that for the densities shown the Thomas-Fermi and Kohn-Sham results agree very closely for
temperature greater than 25 eV.  The Thomas-Fermi model therefore offers a rapid shortcut to realistic potentials
for sufficiently high temperatures.  
\begin{figure}
\begin{center}
\includegraphics[scale=0.3]{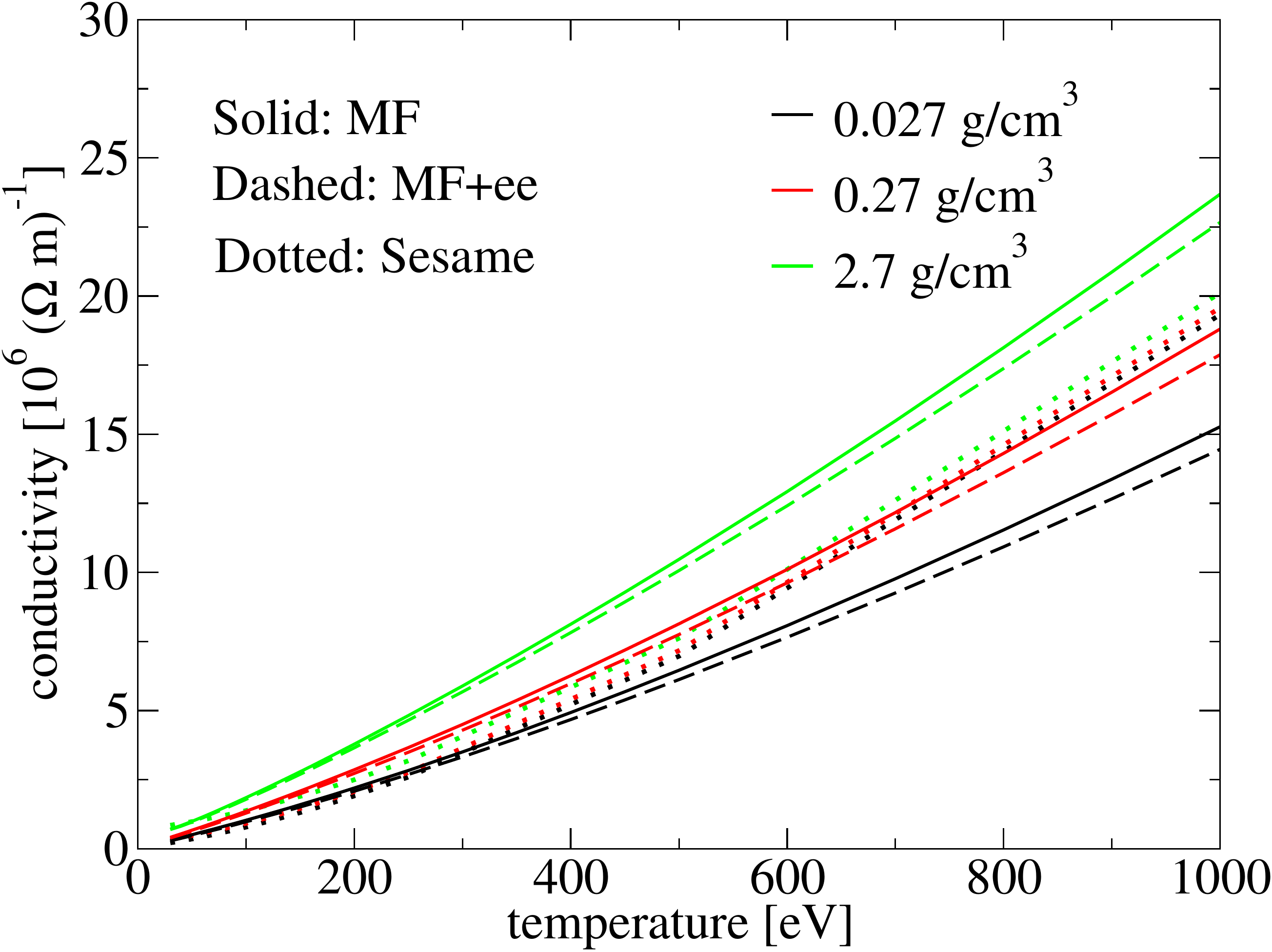}
\end{center}
\caption{(Color online) 
Electrical conductivity of aluminum using the $V^{MF}(r)$ potential
compared to Sesame 29371 \cite{desjarlais01,lee84}.  The mean force
calculations suggest that conductivity is significantly more sensitive to density than
the Sesame model indicates.
}
\label{fig_san}
\end{figure}

In figure \ref{fig_san} we compare calculations based on the present $V^{MF}(r)$ model to the Sesame table 29371, based on the
model of \cite{desjarlais01,lee84}, up to 1000 eV.  We find that the present model shows much more sensitivity to density 
than Sesame 29371 does and that the effect of electron-electron collisions remains relatively small even at these
high temperatures.  At 1000 eV the present model is $\sim$ 13 \% larger for 2.7 g/cm$^3$, while at 0.027 g/cm$^3$,
the present model gives conductivities $\sim$ 25 \% lower.

\section{conclusions}
We have presented a potential of mean force that, coupled with the relaxation time approximation, is used to calculate
electrical conductivity of dense plasmas.  Compared to previously used potentials we found improved agreement
with experiments and bench mark calculations across a range of temperatures and densities for aluminum.  
We note that the new potential takes into account the self-consistent ionic structure factor implicitly, in contrast
to other methods where the structure factor appears explicitly.  This is important because to generate the potential
of mean force, one needs a model that self-consistently solves the quantum Ornstein Zernike equations, whereas previously
the scattering potential and structure factor have often been generated by distinct models.

In applying the new model the influence of electron-electron collisions was included using the fit formula of reference \cite{reinholz15}.
It was found that for dense hydrogen this was essential for agreement with other methods that are thought to be
accurate, indicating the importance and correctness of their inclusion.  In comparison with experiments on aluminum
by DeSilva {\it et al} \cite{desilva98}, it was found that inclusion of electron-electron collisions significantly improved
agreement with both the experiments and bench mark calculations.

We also tested the effect of generating the potential of mean force with a Thomas-Fermi DFT based model as opposed to the more
accurate Kohn-Sham DFT calculation.  It was found that for aluminum from 0.027 to 2.7 g/cm$^3$ the TF based model was accurate
for temperatures  greater than $\sim$25 eV, a fact that greatly speeds up the calculations.  It is important to realise
that while the potential can be generated using the TF based model, the conductivity itself is based on a calculation of
the phase shifts in the usual way. 

The new method was compared to Sesame table 29371 for aluminum and was found to have a significantly greater dispersion with respect
to density at high temperature than Sesame.  At low temperatures significant differences (up to a factor of $\sim$ 2) were also observed.

While we have demonstrated that the new potential results in generally improved agreement with bench mark calculations and experiments, 
the model as a whole is not perfect and could be further improved in a number of ways.  Perhaps
the most apparent error is that the potential assumes on average ionic configuration, instead of a weighted range of configurations.  This leads
to structures in the conductivity as a function of density and temperature that are likely to be too pronounced.  This predominately effects
the lower densities and temperatures considered here.  Other potential sources of error or uncertainty include the in influence
of a more realistic density of states factor \cite{starrett16}, or the choice of exchange and correlation potential in the underlying
DFT calculation \cite{ovechkin16}.

We have focused mainly of aluminum, for which most experimental data is available, but there is no fundamental
restriction and the method should be applicable other materials.  Also, it is straightforward to extend to the calculation
of thermal conductivity.  However, recent investigations \cite{desjarlais17} point to the need for care with such calculation with 
respect to the inclusion of electron-electron collisions.  Finally, we point out that this new potential is also
applicable to opacity calculations \cite{shaffer17} where it could aid in the assessment of the influence of
ionic structure on opacity \cite{krief16}.

\section*{Acknowledgments}
The author acknowledges useful conversations with S. Baalrud and thanks T. Sjostrom for providing his DFT-MD data and
M. Desjarlais for useful comments on the manuscript.
This work was performed under the auspices of the United States Department of Energy under contract DE-AC52-06NA25396.

\appendix
\section{Calculation of phase shifts}
Calculation of the scattering amplitudes requires the scattering phase shifts $\eta_l(\epsilon)$ for the potential.
These are calculated in the usual way via continuum state normalization.  However, high plasma temperatures
and low densities require phase shifts at large $l$ ($>10\,000$ are required) and 
$\epsilon$ (up to $\sim$1000 $E_h$).  Such cases are numerically more
challenging and a number of schemes have been proposed to circumvent their calculation \cite{pain10,perrot87}.
We have implemented a different scheme that we found to be accurate and reliable.

The phase shifts are calculated in the semi-classical JWKB approximation with knowledge only
of the scattering potential \cite{handbook}
\begin{eqnarray}
\eta_l(\epsilon) & = & \int\limits_{R_C}^\infty \left[ p^2 - 2 V^{scatt}(r) - \frac{(l+\hf)^2}{r^2} \right]^\hf d\br\nonumber\\
                 &   & - \int\limits_{(l+\hf)/p}^\infty \left[ p^2 - \frac{(l+\hf)^2}{r^2} \right]^\hf d\br \label{cps}
\end{eqnarray}
where $R_C$ is the classical turning point which restricts the integrand of the first integral to be positive energy 
\begin{equation}
 p^2 - 2 V(R_C) - {(l+\hf)^2}{R_C^2}  = 0\label{ctp}
\end{equation}
We have found this calculation to be fast, accurate and stable even for very large l ($>$ 30$\,$000).  Typically
we switch to this calculation for energies greater than 50 $E_h$ or orbital angular momenta greater than 30.

\bibliographystyle{unsrt}
\bibliography{phys_bib}

\begin{thebibliography}{10}

\bibitem{desjarlais02}
M.~P. Desjarlais, J.~D. Kress, and L.~A. Collins.
\newblock Electrical conductivity for warm, dense aluminum plasmas and liquids.
\newblock {\em Phys. Rev. E}, 66:025401, Aug 2002.

\bibitem{sjostrom15}
Travis Sjostrom and J\'er\^ome Daligault.
\newblock Ionic and electronic transport properties in dense plasmas by
  orbital-free density functional theory.
\newblock {\em Phys. Rev. E}, 92:063304, 2015.

\bibitem{holst11}
Bastian Holst, Martin French, and Ronald Redmer.
\newblock Electronic transport coefficients from \textit{ab initio} simulations
  and application to dense liquid hydrogen.
\newblock {\em Phys. Rev. B}, 83:235120, Jun 2011.

\bibitem{french14}
Martin French and Thomas~R Mattsson.
\newblock Thermoelectric transport properties of molybdenum from ab initio
  simulations.
\newblock {\em Physical Review B}, 90(16):165113, 2014.

\bibitem{hu14}
S.~X. Hu, L.~A. Collins, T.~R. Boehly, J.~D. Kress, V.~N. Goncharov, and
  S.~Skupsky.
\newblock First-principles thermal conductivity of warm-dense deuterium plasmas
  for inertial confinement fusion applications.
\newblock {\em Phys. Rev. E}, 89:043105, Apr 2014.

\bibitem{desjarlais17}
Michael~P. Desjarlais, Christian~R. Scullard, Lorin~X. Benedict, Heather~D.
  Whitley, and Ronald Redmer.
\newblock Density-functional calculations of transport properties in the
  nondegenerate limit and the role of electron-electron scattering.
\newblock {\em Phys. Rev. E}, 95:033203, Mar 2017.

\bibitem{sperling15}
P.~Sperling, E.~J. Gamboa, H.~J. Lee, H.~K. Chung, E.~Galtier, Y.~Omarbakiyeva,
  H.~Reinholz, G.~R\"opke, U.~Zastrau, J.~Hastings, L.~B. Fletcher, and S.~H.
  Glenzer.
\newblock Free-electron x-ray laser measurements of collisional-damped plasmons
  in isochorically heated warm dense matter.
\newblock {\em Phys. Rev. Lett.}, 115:115001, 2015.

\bibitem{desilva98}
A.~W. DeSilva and J.~D. Katsouros.
\newblock Electrical conductivity of dense copper and aluminum plasmas.
\newblock {\em Phys. Rev. E}, 57:5945--5951, May 1998.

\bibitem{clerouin08}
Jean Cl{\'e}rouin, Pierre Noiret, Victor~N. Korobenko, and Anatoly~D. Rakhel.
\newblock Direct measurements and ab initio simulations for expanded fluid
  aluminum in the metal-nonmetal transition range.
\newblock {\em Physical Review B}, 78(22):224203, 2008.

\bibitem{clerouin12}
J.~Cl{\'e}rouin, P.~Noiret, P.~Blottiau, V.~Recoules, B.~Siberchicot,
  P.~Renaudin, C.~Blancard, G.~Faussurier, B.~Holst, and C.~E. Starrett.
\newblock A database for equations of state and resistivities measurements in
  the warm dense matter regime.
\newblock {\em Physics of Plasmas}, 19(8):082702, 2012.

\bibitem{milchberg88}
H.~M. Milchberg, R.~R. Freeman, S.~C. Davey, and R.~M. More.
\newblock Resistivity of a simple metal from room temperature to ${10}^{6}$ k.
\newblock {\em Phys. Rev. Lett.}, 61:2364--2367, Nov 1988.

\bibitem{benage99}
J.~F. Benage, W.~R. Shanahan, and M.~S. Murillo.
\newblock Electrical resistivity measurements of hot dense aluminum.
\newblock {\em Phys. Rev. Lett.}, 83:2953--2956, Oct 1999.

\bibitem{lee84}
Yim~T. Lee and R.~M. More.
\newblock An electron conductivity model for dense plasmas.
\newblock {\em The Physics of fluids}, 27(5):1273--1286, 1984.

\bibitem{burrill16}
D.J. Burrill, D.V. Feinblum, M.R.J. Charest, and C.E. Starrett.
\newblock Comparison of electron transport calculations in warm dense matter
  using the ziman formula.
\newblock {\em High Energy Density Physics}, 19:1 -- 10, 2016.

\bibitem{ziman61}
J.M. Ziman.
\newblock A theory of the electrical properties of liquid metals. i: The
  monovalent metals.
\newblock {\em Philosophical Magazine}, 6(68):1013--1034, 1961.

\bibitem{rinker88}
George~A. Rinker.
\newblock Systematic calculations of plasma transport coefficients for the
  periodic table.
\newblock {\em Phys. Rev. A}, 37:1284--1297, 1988.

\bibitem{ovechkin16}
A.A. Ovechkin, P.A. Loboda, and A.L. Falkov.
\newblock Transport and dielectric properties of dense ionized matter from the
  average-atom reseos model.
\newblock {\em High Energy Density Physics}, 20:38 -- 54, 2016.

\bibitem{perrot87}
Fran\c{c}ois Perrot and M.~W.~C. Dharma-wardana.
\newblock Electrical resistivity of hot dense plasmas.
\newblock {\em Phys. Rev. A}, 36:238--246, Jul 1987.

\bibitem{scaalp}
G\'erald Faussurier, Christophe Blancard, Philippe Coss\'e, and Patrick
  Renaudin.
\newblock Equation of state, transport coefficients, and stopping power of
  dense plasmas from the average-atom model self-consistent approach for
  astrophysical and laboratory plasmas.
\newblock {\em Physics of Plasmas}, 17(5), 2010.

\bibitem{sterne07}
P.A. Sterne, S.B. Hansen, B.G. Wilson, and W.A. Isaacs.
\newblock Equation of state, occupation probabilities and conductivities in the
  average atom purgatorio code.
\newblock {\em High Energy Density Physics}, 3(1–2):278 -- 282, 2007.
\newblock Radiative Properties of Hot Dense Matter.

\bibitem{starrett12a}
C.~E. Starrett, J~Cl{\'e}rouin, V~Recoules, J.~D. Kress, L.~A. Collins, and
  D.~E. Hanson.
\newblock Average atom transport properties for pure and mixed species in the
  hot and warm dense matter regimes.
\newblock {\em Physics of Plasmas (1994-present)}, 19(10):102709, 2012.

\bibitem{perrot99}
Fran{\c{c}}ois Perrot and M.W.C. Dharma-Wardana.
\newblock Theoretical issues in the calculation of the electrical resistivity
  of plasmas.
\newblock {\em International journal of thermophysics}, 20(4):1299--1311, 1999.

\bibitem{starrett16b}
Charles~Edward Starrett.
\newblock Kubo--greenwood approach to conductivity in dense plasmas with
  average atom models.
\newblock {\em High Energy Density Physics}, 19:58--64, 2016.

\bibitem{baalrud13}
Scott~D. Baalrud and J\'er\^ome Daligault.
\newblock Effective potential theory for transport coefficients across coupling
  regimes.
\newblock {\em Phys. Rev. Lett.}, 110:235001, Jun 2013.

\bibitem{chihara91}
Junzo Chihara.
\newblock Unified description of metallic and neutral liquids and plasmas.
\newblock {\em Journal of Physics: Condensed Matter}, 3(44):8715, 1991.

\bibitem{reinholz15}
H.~Reinholz, G.~R{\"o}pke, S.~Rosmej, and R.~Redmer.
\newblock Conductivity of warm dense matter including electron-electron
  collisions.
\newblock {\em Physical Review E}, 91(4):043105, 2015.

\bibitem{pain10}
J.C. Pain and G.~Dejonghe.
\newblock Electrical resistivity in warm dense plasmas beyond the average-atom
  model.
\newblock {\em Contributions to Plasma Physics}, 50(1):39--45, 2010.

\bibitem{faussurier15}
G\'erald Faussurier and Christophe Blancard.
\newblock Resistivity saturation in warm dense matter.
\newblock {\em Phys. Rev. E}, 91:013105, Jan 2015.

\bibitem{rozsnyai08}
Balazs~F. Rozsnyai.
\newblock Electron scattering in hot/warm plasmas.
\newblock {\em High Energy Density Physics}, 4(1):64--72, 2008.

\bibitem{evans73}
R.~Evans, B.~L. Gyorfey, N.~Szabo, and J.~M. Ziman.
\newblock On the resistivity of liquid transition metals.
\newblock In S.~Takeuchi, editor, {\em The properties of liquid metals}. Taylor
  and Francis, London, 1973.

\bibitem{daligault16}
J{\'e}r{\^o}me Daligault, Scott~D. Baalrud, Charles~E. Starrett, Didier Saumon,
  and Travis Sjostrom.
\newblock Ionic transport coefficients of dense plasmas without molecular
  dynamics.
\newblock {\em Physical review letters}, 116(7):075002, 2016.

\bibitem{starrett14b}
C.~E. Starrett, D.~Saumon, J.~Daligault, and S.~Hamel.
\newblock Integral equation model for warm and hot dense mixtures.
\newblock {\em Physical Review E}, 90(3):033110, 2014.

\bibitem{starrett13}
C.~E. Starrett and D.~Saumon.
\newblock Electronic and ionic structures of warm and hot dense matter.
\newblock {\em Phys. Rev. E}, 87:013104, Jan 2013.

\bibitem{starrett14}
C.E. Starrett and D.~Saumon.
\newblock A simple method for determining the ionic structure of warm dense
  matter.
\newblock {\em High Energy Density Physics}, 10:35 -- 42, 2014.

\bibitem{desjarlais01}
M.P. Desjarlais.
\newblock Practical improvements to the lee-more conductivity near the
  metal-insulator transition.
\newblock {\em Contributions to Plasma Physics}, 41(2-3):267--270, 2001.

\bibitem{starrett16}
C.~E. Starrett and D.~Saumon.
\newblock Equation of state of dense plasmas with pseudoatom molecular
  dynamics.
\newblock {\em Phys. Rev. E}, 93:063206, Jun 2016.

\bibitem{starrett15b}
C.~E. Starrett and D.~Saumon.
\newblock Models of the elastic x-ray scattering feature for warm dense matter.
\newblock {\em Phys. Rev. E}, 92:033101, 2015.

\bibitem{potekhin96}
A.~Yu Potekhin and D.G. Yakovlev.
\newblock Electron conduction along quantizing magnetic fields in neutron star
  crusts. ii. practical formulae.
\newblock {\em Astronomy and Astrophysics}, 314:341--352, 1996.

\bibitem{kuhlbrodt01}
S.~Kuhlbrodt, R.~Redmer, A.~Kemp, and J.~Meyer-ter Vehn.
\newblock Conductivities in hot aluminium plasma.
\newblock {\em Contributions to Plasma Physics}, 41(1):3--14, 2001.

\bibitem{wardana06}
M.W.C. Dharma-Wardana.
\newblock Static and dynamic conductivity of warm dense matter within a
  density-functional approach: Application to aluminum and gold.
\newblock {\em Physical Review E}, 73(3):036401, 2006.

\bibitem{sjostrom_p}
T.~Sjostrom.
\newblock {\em Private communication}.

\bibitem{dwp92}
M.W.C. Dharma-Wardana and Fran{\c{c}}ois Perrot.
\newblock Resistivity and dynamic conductivity of laser-pulse heated aluminum
  up to 106 k and along the shock hugoniot up to 20 mbars.
\newblock {\em Physics Letters A}, 163(3):223--227, 1992.

\bibitem{shaffer17}
Nathaniel~R. Shaffer, Natalie~G. Ferris, J.~Colgan, David~Parker Kilcrease, and
  Charles~Edward Starrett.
\newblock Free-free opacity in dense plasmas with an average atom model.
\newblock {\em High Energy Density Physics}, 23:31--37, 2017.

\bibitem{krief16}
Menahem Krief, Yair Kurzweil, Alexander Feigel, and Doron Gazit.
\newblock The effect of ionic correlations on radiative properties in the solar
  interior and terrestrial experiments.
\newblock {\em arXiv preprint arXiv:1611.09339}, 2016.

\bibitem{handbook}
G.~W. F.~Drake (Editor).
\newblock {\em Springer handbook of atomic, molecular and optical physcis}.
\newblock Springer, 2005.

\end{thebibliography}

\end{document}